\documentclass[12pt]{article}

\usepackage{epsfig}
\usepackage{graphicx}

\begin{document}

\begin{center}
{\bfseries $\rho$ meson production in ultraperipheral dAu collision}

\vskip 5mm

S.L. Timoshenko$^{1 \dag}$ (for STAR collaboration)

\vskip 5mm

{\small
(1) {\it
Moscow Engineering Physical Institute (State University), Moscow, Russia
}
$\dag$ {\it
E-mail: tim@intphys3.mephi.ru
}}
\end{center}

\vskip 5mm

\begin{center}
\begin{minipage}{150mm}
\centerline{\bf Abstract}
 Ultra-peripheral heavy ion collisions involve long range electromagnetic interactions at impact parameters larger than twice the nuclear radius, where no nucleon-nucleon collisions occur. We report on the first observation of $\rho$ meson  production in $dAu\to dAu\rho^0$ and $dAu\to npAu\rho^0$.
\end{minipage}
\end{center}

\vskip 10mm

In ultraperipheral heavy-ion collisions, two nuclei geometrically ``miss'' each other. The impact parameter  is larger than twice nuclear radius. In such conditions hadronic interactions do not occur\cite{bib1}. Relativistic ions are sources of electromagnetic and pomeron fields. The electromagnetic fields have a large radius of interaction, than strong interactions. So, in ultraperipheral collisions, the nuclei interact by two photons or photon-pomeron exchange. The photon flux is proportional to the square of the nuclear charge $Z^{2}$\cite{bib2}, and the forward cross section for elastic $\rho^{0}A$ scattering scales as $A^{4/3}$ for surface coupling and $A^{2}$ in the bulk. Thus the cross section for photon-pomeron interaction $\sim A^{2}Z^{2}$  for ``heavy'' states (as like J/psi) and $Z^{2}A^{4/3}$ for lighter mesons ($\rho,\omega,\phi$).\par

%% To insert figure (with the help of epsf.sty)
\begin{figure}[htbp]
\begin{center}\mbox{\epsfig{file=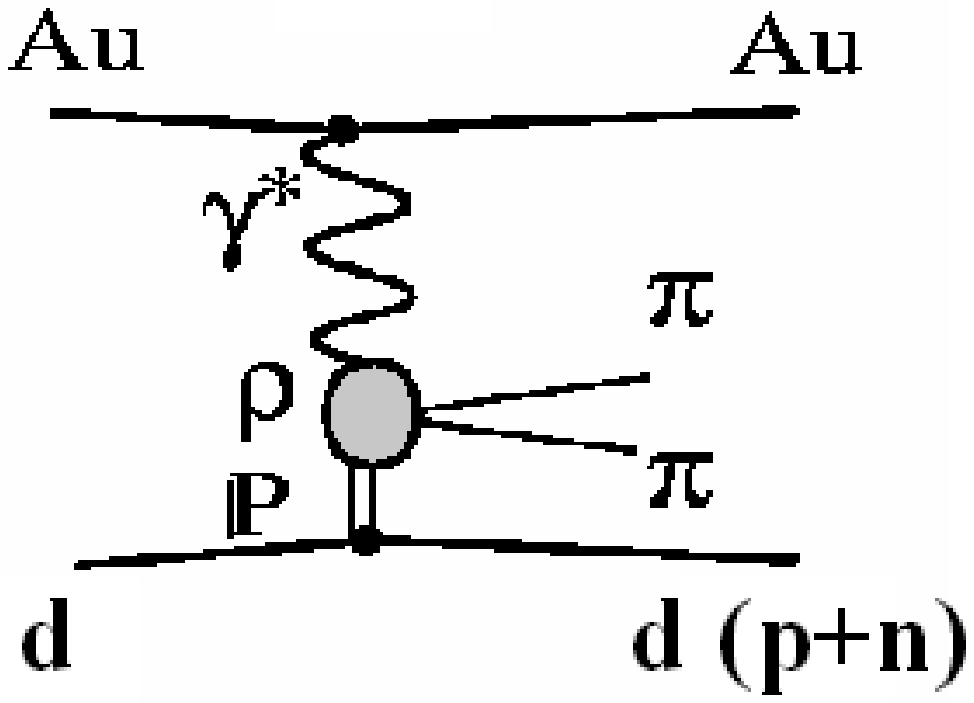,height=50mm,width=60mm} \quad\quad\epsfig{file=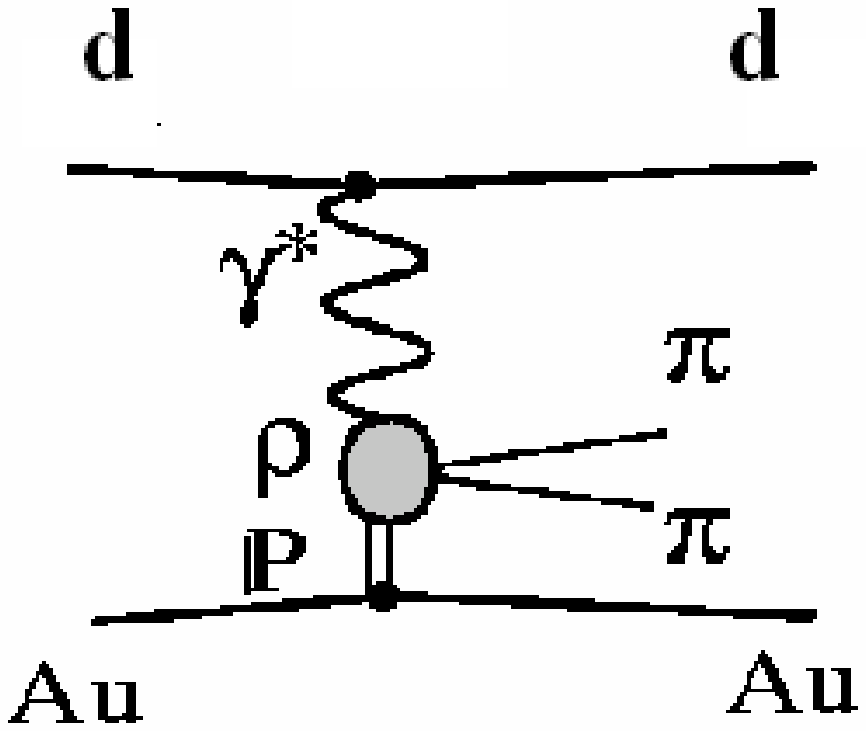,height=50mm,width=60mm}}\end{center} 
\begin{center}\quad\quad\ Fig.1(a)\quad\quad\quad\quad\quad\quad\quad\quad\quad\quad\quad\quad Fig.1(b) \end{center}
\end{figure}

Exclusive vector meson production $dAu\to d(np)Au\rho^{0}$ (fig.1(a,b)) can be described by using Weizsacker-Williams approach\cite{bib2} to photon and pomeron fluxes and vector dominance model.  Photon emitted by one nucleus can be consider as a state of virtual photons plus some fluctuations of quark-antiquark pairs. When the nucleus absorbs ``photonic'' part of wave function, the quark-antiquark pairs contribution becomes dominant. This pair can elastically scatters on the other nucleus and appears as a real vector meson.\par

In ultraperipheral deuteron gold interactions there are  two mechanisms of $\rho$-meson production. First, a photon emitted  by the gold interacts with the deuteron producing $\rho$-meson and in this case deuteron can break up ($\gamma d\to np\rho^{0}$) or remains in the ground state($\gamma d\to d\rho^{0}$). The cross section of such process is approximately a factor of 16,000 ($Z^{2}A^{4/3}\sim79^{2}2^{4/3}$). But there is a process when a photon emitted by the deuteron interacts with the gold. The cross section of such process is approximately a factor of 1,200 ($Z^{2}A^{4/3}\sim1^{2}197^{4/3}$). One can see that the cross section for the first reactions is much large than that for the second. One therefore, at experiment will dominate production $\rho$-meson in photon-deuteron interaction.

In the year 2002 RHIC at Brookhaven National Laboratory gold and deuteron nucleus collided at$\sqrt{s_{NN}}=200 $GeV. In the Solenoidal Tracker at RHIC (STAR)\cite{bib4}, charge particles are reconstructed in a 4.2 meter long, 4 meter diameter a cylindrical time projection chamber (TPC). A solenoid magnet surrounds TPC\cite{bib5}. In 2002 the TPC was operated in a 0.5 T solenoidal magnetic field.  Particles were identified by their energy loss in the TPC. The TPC is surrounded by a cylindrical central trigger barrel (CTB). CTB consist of 240 scintillator slats covering$\mid\eta\mid<1$ . For registration of neutrons there are two zero degree calorimeters (ZDC) at 18 m from the interaction point\cite{bib6}. These calorimeters are sensitive to single neutrons and have efficiency of close to 100 percent. 

%% To insert figure (with the help of epsf.sty)
\begin{figure}[htbp]
\begin{center}\mbox{\epsfig{file=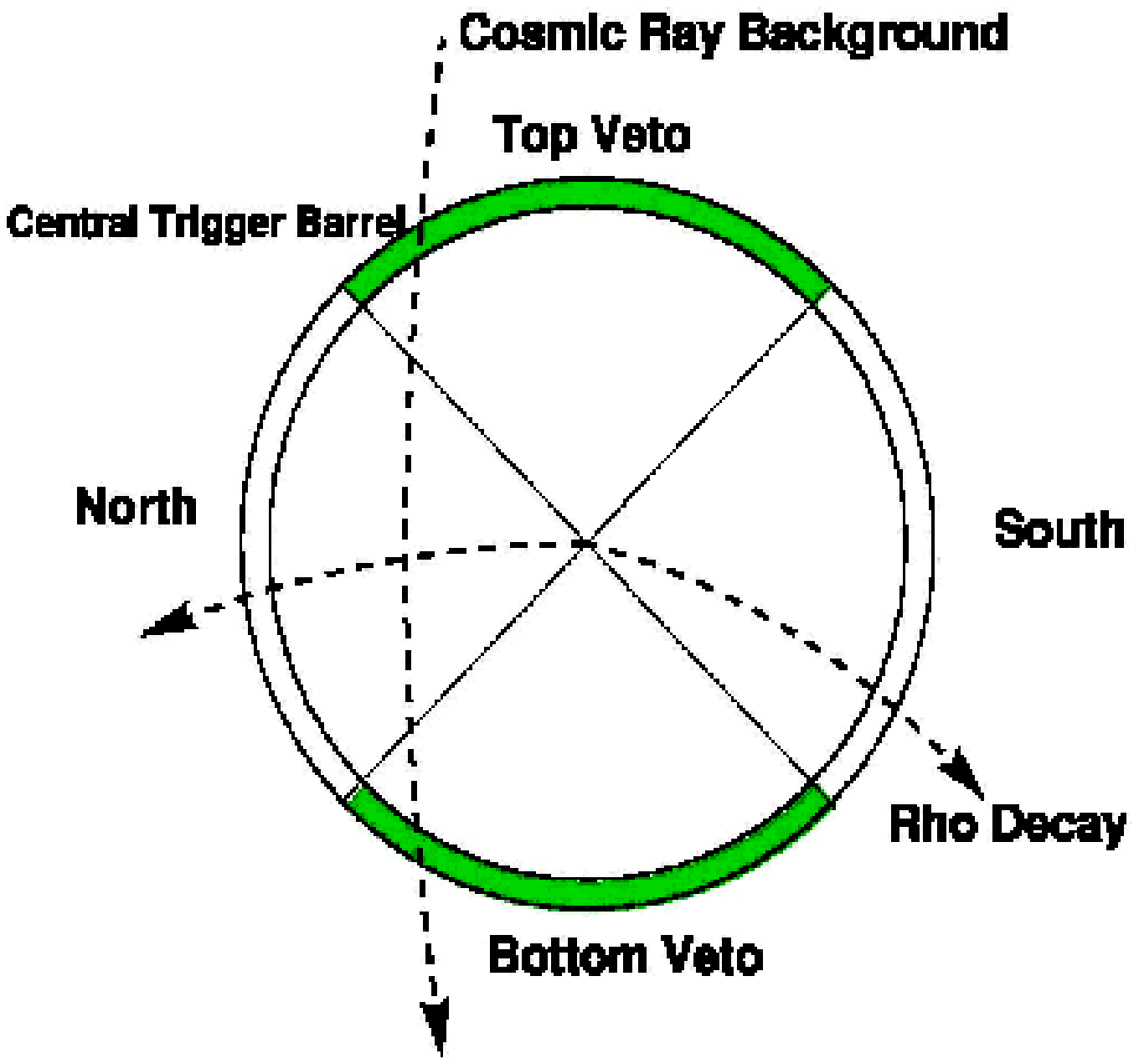,height=60mm,width=80mm}}\end{center}
\begin{center} Fig.2 \end{center}
\end{figure}
Exclusive $\rho^0$ production in UPC has a distinctive experimental signature: the $\pi^{+}\pi^{-}$ decay products of the $\rho^0$ meson are observed in an  otherwise ``empty'' spectrometer. Two different triggers were used for this analysis. For $dAu\to dAu\rho^{0}$, about 700,000  events were collected using a low-multiplicity ``topology'' trigger. The topology trigger was designed to detect the products $\rho^0$ decay in the CTB system. The CTB was divided into four azimuthal quadrants (fig.2). Selected events were required to have at least one hit in the north and south sectors. The top and bottom quadrants were used as a veto, to reject cosmic rays. To study $dAu\to npAu\rho^{0}$ reaction, about 250,000 thousands events are used for the analyses another trigger consists from ``topology'' trigger and ZDC(West). This trigger (``topology-ZDC'') required the detection of a neutron the deuteron breakup.  The main background for two triggers are cosmic rays, beam gas interaction, pile-up. \par

For our analysis we selected events with exactly two reconstructed tracks in the TPC. Total charge of the two tracks must be equal to zero. Number of hits in the track can be more then 13. Events were accepted if two tracks were consistent from a single vertex.  \par

%% To insert figure (with the help of epsf.sty)
\begin{figure}[htbp]
\begin{center}\mbox{\epsfig{file=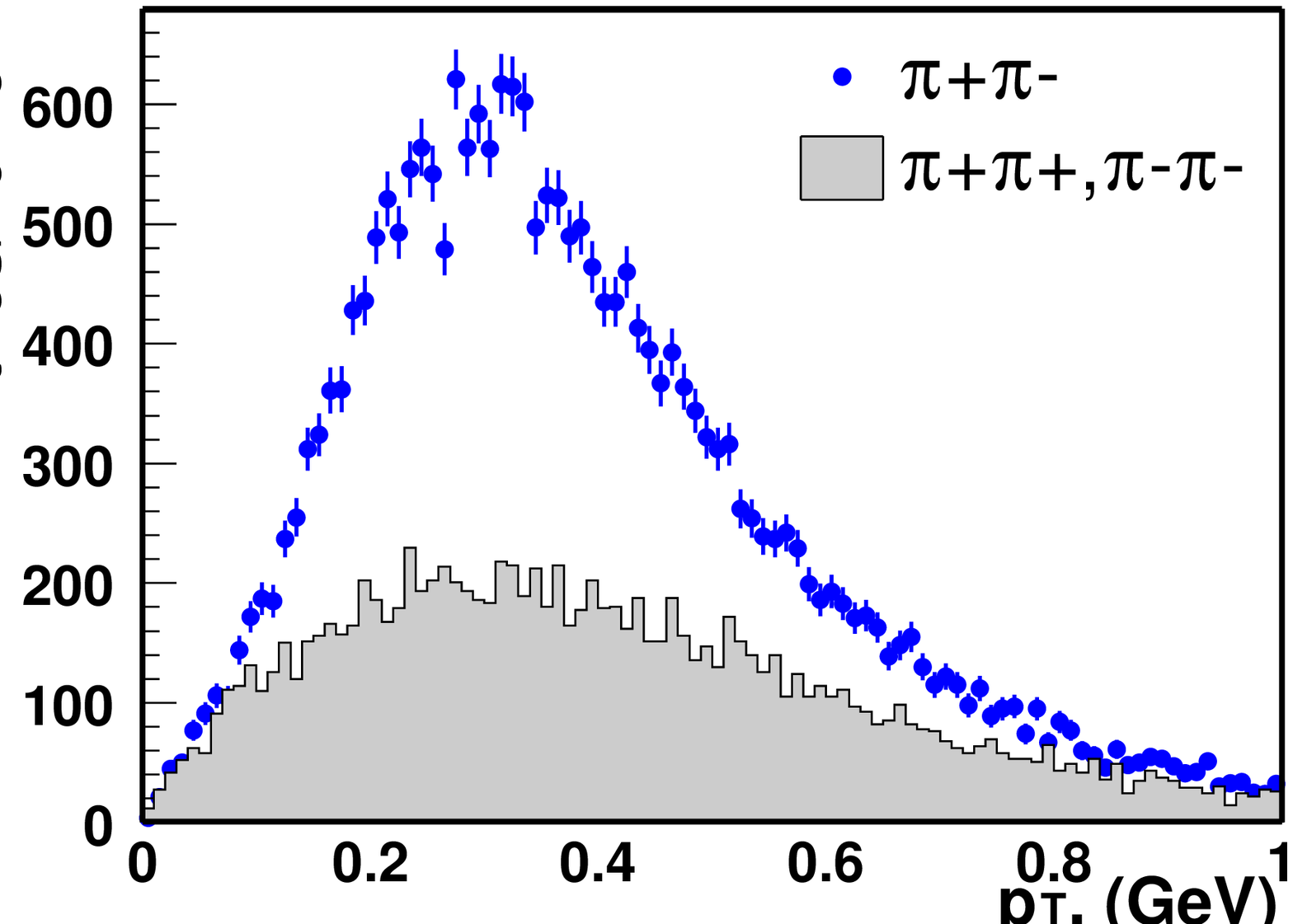,height=50mm,width=70mm}\quad\quad\epsfig{file=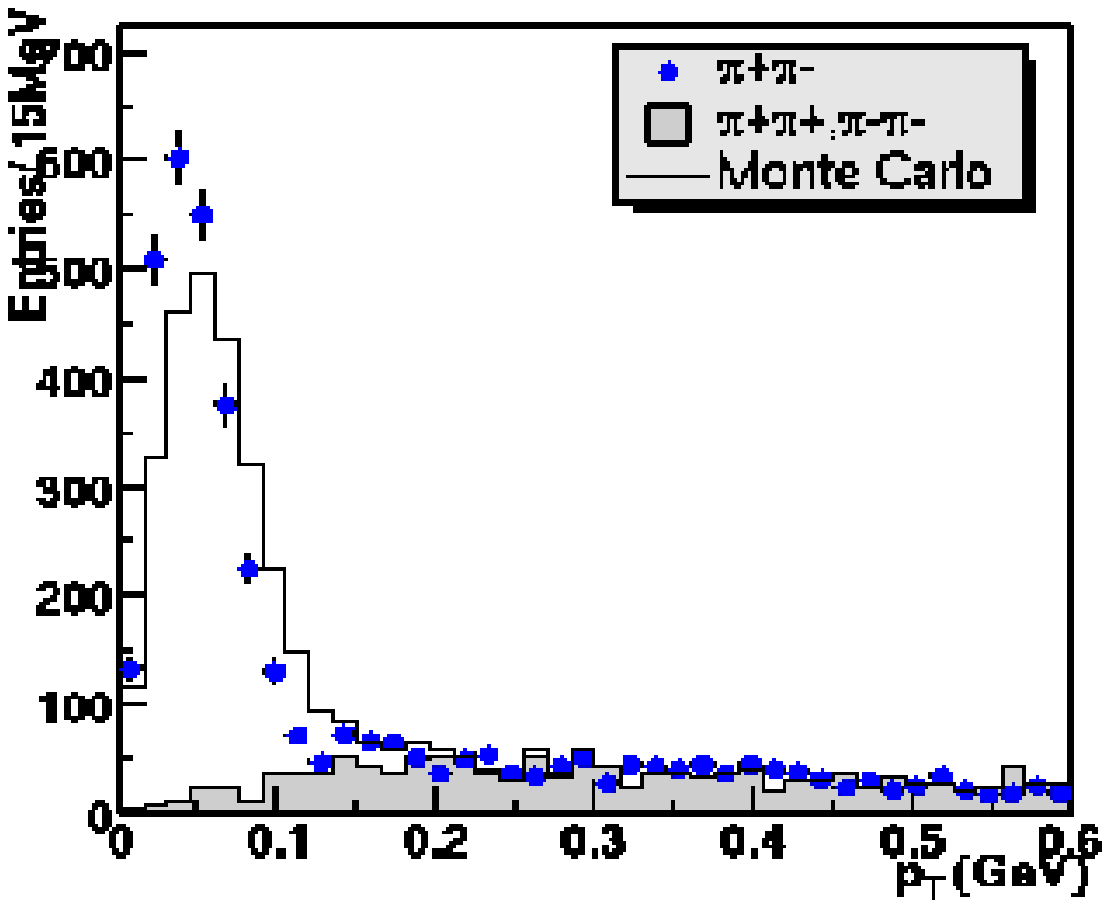,height=50mm,width=70mm}}\end{center}
\begin{center}\quad Fig.3\quad\quad\quad\quad\quad\quad\quad\quad\quad\quad\quad\quad\quad\quad\quad Fig.4 \end{center}
\end{figure}
Figure 3 shows the transverse momentum spectrum of oppositely charged pion pairs production in deuteron gold collisions. As one can see, there is a large peak about 300 MeV. It is necessary to remind $p_{T}$ spectrum of $\rho^0$  mesons produced in gold-gold collision\cite{bib7}. Figure 4 shows the transverse momentum spectrum for $\pi^{+}\pi^{-}$ pairs. A clear peak, the signature for coherent coupling, can be observed at $p_{T}<$ 100 MeV . This is consistent with coherent $\rho^0$meson production. A background model from like-sign combination pairs (shaded histogram), which is normalized to the signal at $p_{T}$$>$250 MeV, does not show such a peak. But in $dAu$ collisions a background couldn't normalize to the signal. \par

%% To insert figure (with the help of epsf.sty)
\begin{figure}[htbp]
\begin{center}\mbox{\epsfig{file=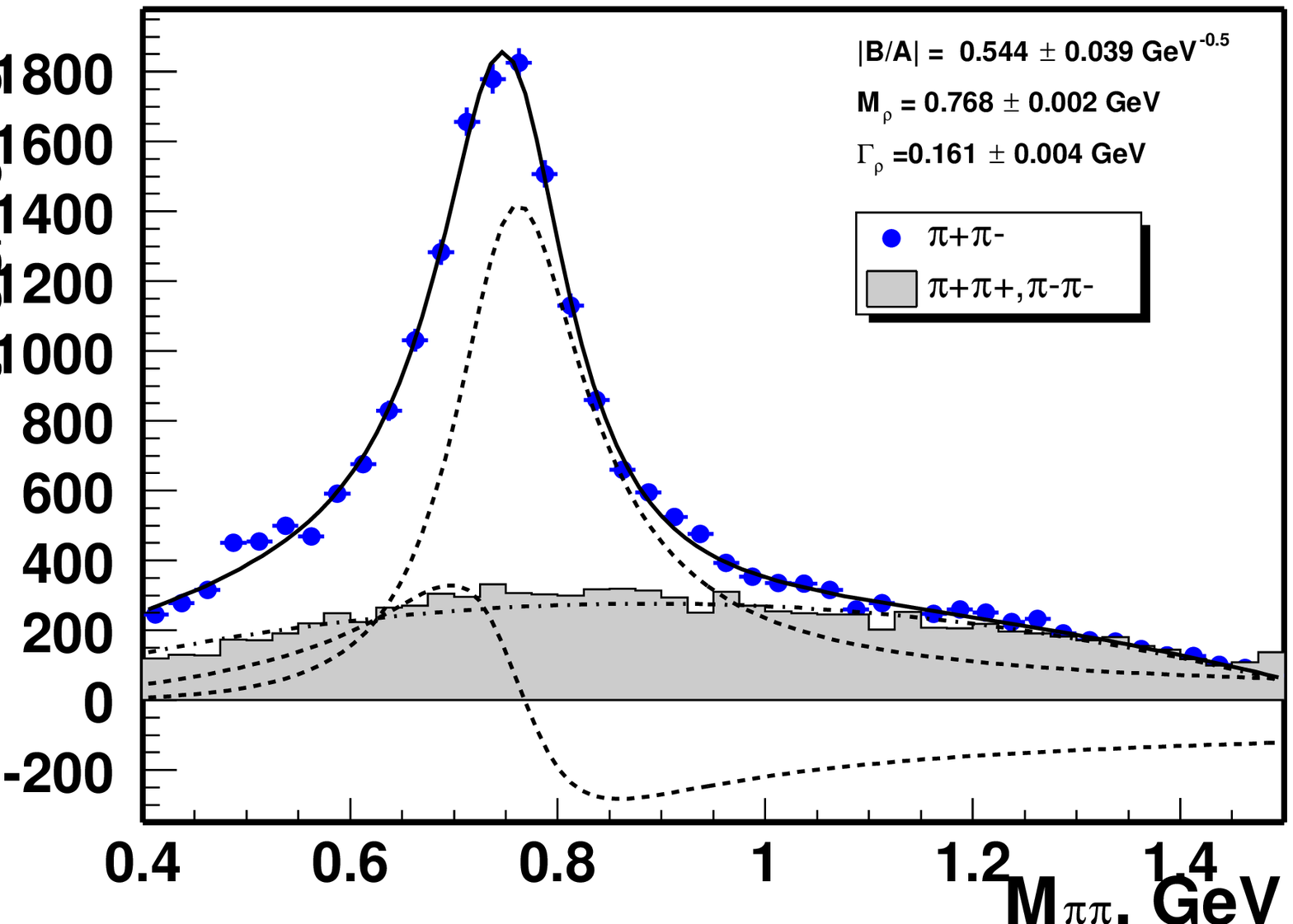,height=50mm,width=70mm}\quad\quad\epsfig{file=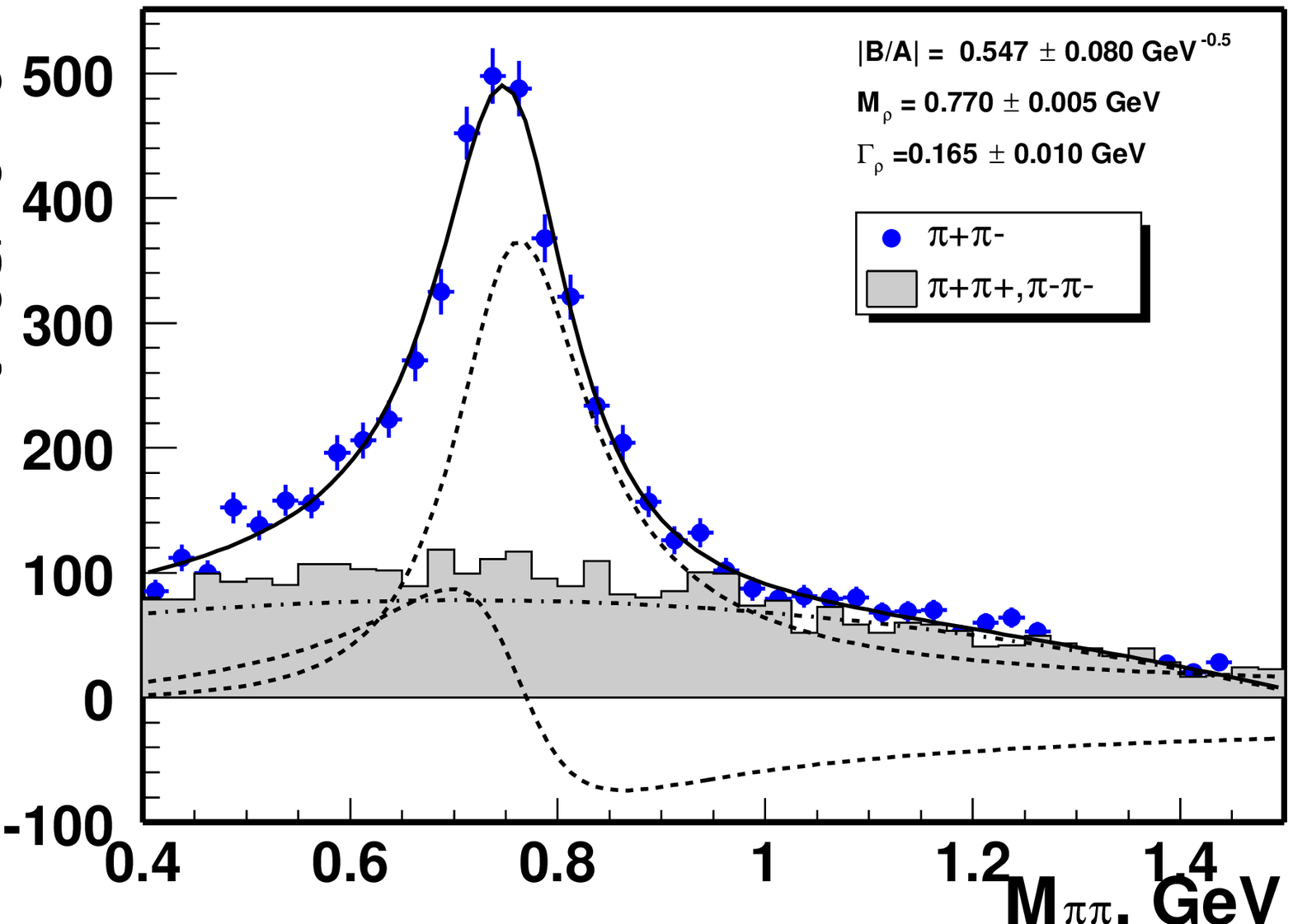,height=50mm,width=70mm}}\end{center}
\begin{center}\quad Fig.5(a)\quad\quad\quad\quad\quad\quad\quad\quad\quad\quad\quad\quad\quad\quad\quad Fig.5(b) \end{center}
\end{figure}
Figure 5(a,b) shows invariant mass of the  $\pi^{+}\pi^{-}$  pairs for ``topology'' and ``topology-ZDC'' triggers. There is a clear $\rho^0$ peak. About 1,500 and 14,000 events around the $\rho^0$ mass from the ``topology'' and ``topology-ZDC''  triggers respectively were used in the analysis.The fit (solid) is the sum of a relativistic Breit-Wigner for $\rho^0$  production and a Soding interference term for direct $\pi^{+}\pi^{-}$   production (both dashed)\cite{bib8}. A second order polynomial (dash-dotted) describes the combinatorial background (shaded histogram). The interference shifts the of the distribution to lower masses $M_{\pi^{+}\pi^{-}}$. Rho mass and width are consistent with its values from Particle date book. The direct $\pi\pi$ to  ratio agrees with ZEUS collaboration in $\gamma p$ interactions\cite{bib9}. \par

We study the $p_{T}$ spectra using the variable $t_{\perp}=p^{2}_{T}$. At RHIC energies, the longitudinal component of the 4-momentum transfer is small, so $t\approx t_{\perp}$. Figure 6(a,b) shows t spectrum for two different triggers. One can see that t spectrum at fig 6(a) decrease at small t. Our data we have compared with the data from fixed targed (fig.7)\cite{bib10} $\gamma d$ experiment\cite{bib11}. One can see the  incoherent behavior of distribution similar to t distribution at fig.6(a). One can conclude that $\rho$ meson production in reaction when a photon interaction with deuteron and  then deuteron break up occur is a incoherent reaction. While t spectrum (fig.6(b)) is similar to coherent behavior. The data were fitted with the function $dN/dt\sim exp(-bt)$.  We obtained $b=9.2\pm0.2 GeV^{-2}$ and $b=8.4\pm0.4 GeV^{-2}$ (statistical errors only) accordingly for  ``topology'' and ``topology-ZDC''  triggers. \par

%% To insert figure (with the help of epsf.sty)
\begin{figure}[htbp]
\begin{center}\mbox{\epsfig{file=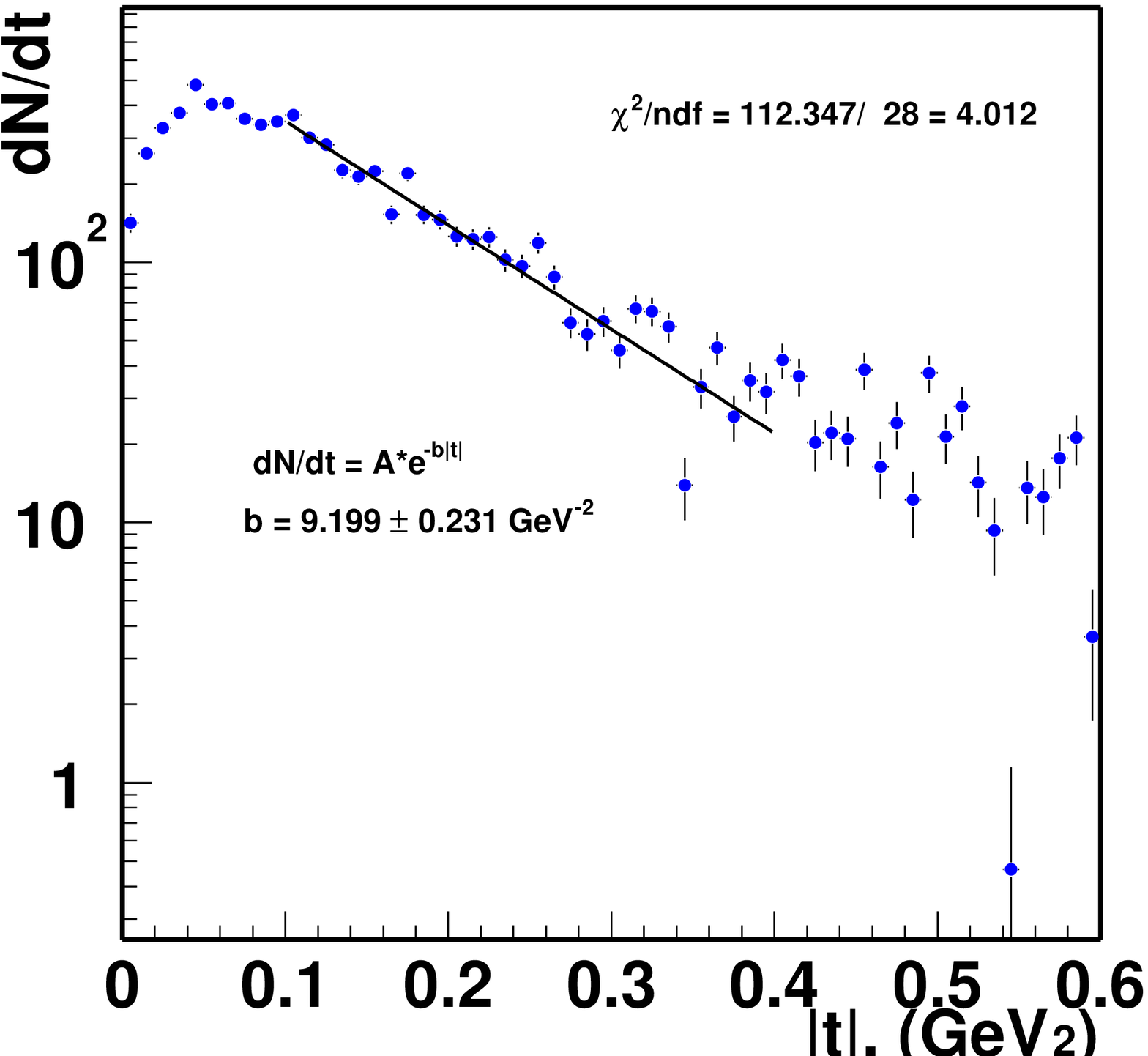,height=50mm,width=50mm}\quad\quad\epsfig{file=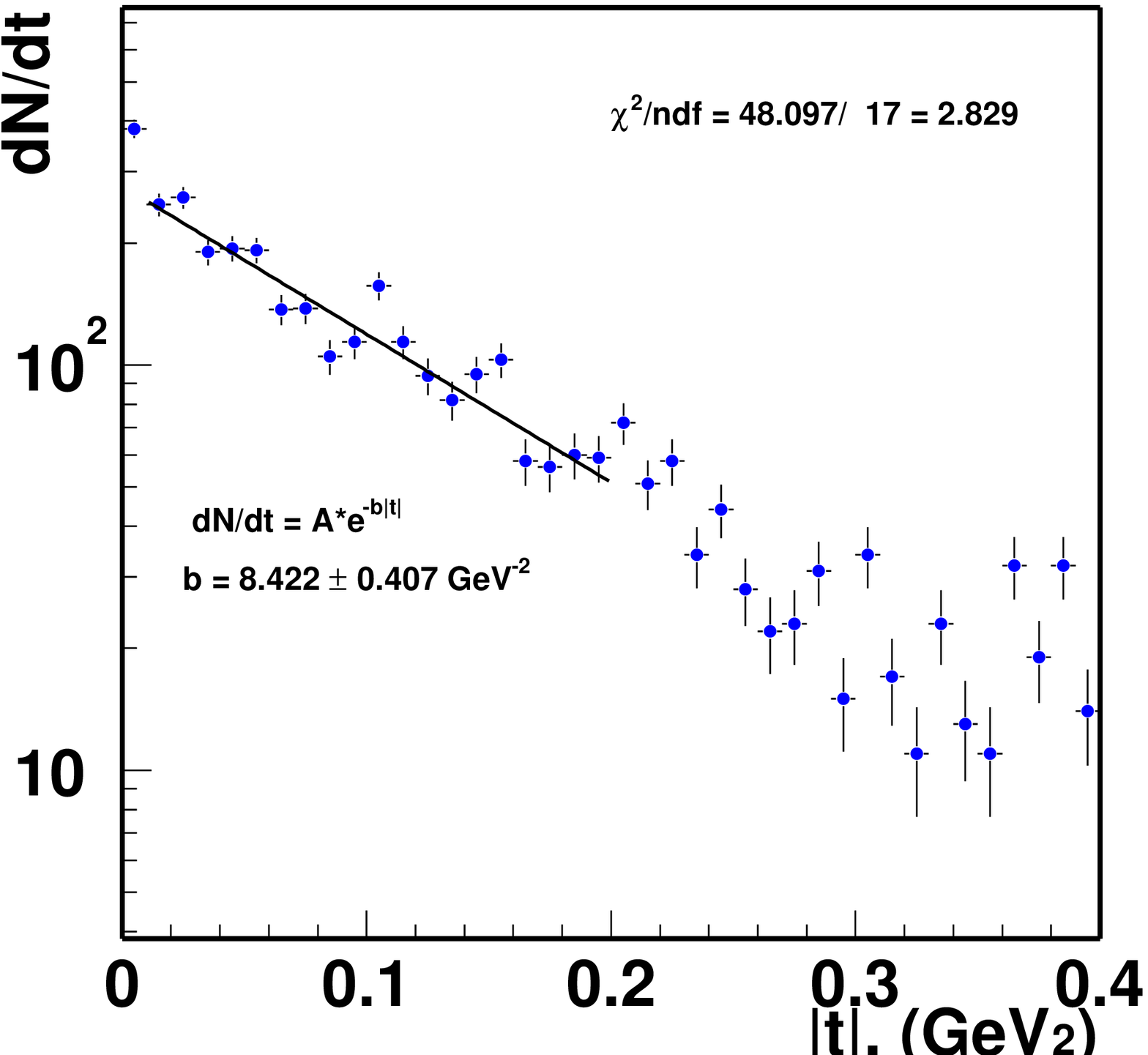,height=50mm,width=50mm}\quad\epsfig{file=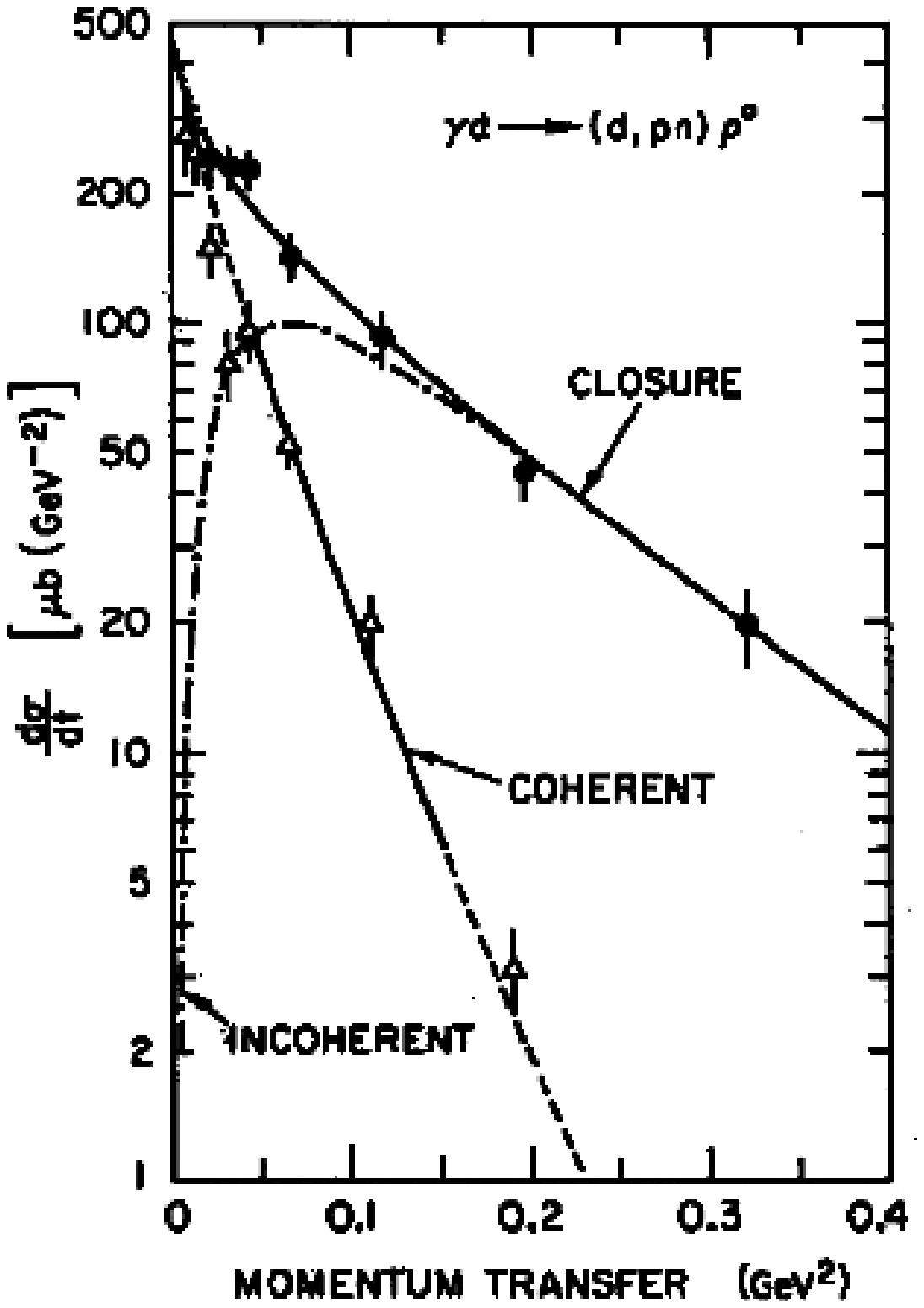,height=50mm,width=50mm}}\end{center}
\begin{center}Fig.6(a)\quad\quad\quad\quad\quad\quad\quad\quad\quad\quad\quad Fig.6(b)\quad\quad\quad\quad\quad\quad\quad\quad\quad\quad\quad Fig.7 \end{center}
\end{figure}

In summary, the first measurements of $\rho^0$  production in $dAu\to dAu\rho^0$ and $dAu\to npAu\rho^0$, confirm the existence of vector meson production in ultra-peripheral heavy ion collisions. It was shown, what in reaction with deuteron break up the $\rho$-meson production mechanism is incoherent. \par

\end{document}